# Improving the Intelligibility of Electric and Acoustic Stimulation Speech Using Fully Convolutional Networks Based Speech Enhancement

Natalie Yu-Hsien Wang, Hsiao-Lan Sharon Wang, Tao-Wei Wang, Szu-Wei Fu, Xugan Lu, Yu Tsao, and Hsin-Min Wang

*Abstract*—The combined electric and acoustic stimulation (EAS) has demonstrated better speech recognition than conventional cochlear implant (CI) and yielded satisfactory performance under quiet conditions. However, when noise signals are involved, both the electric signal and the acoustic signal may be distorted, thereby resulting in poor recognition performance. To suppress noise effects, speech enhancement (SE) is a necessary unit in EAS devices. Recently, a time-domain speech enhancement algorithm based on the fully convolutional neural networks (FCN) with a short-time objective intelligibility (STOI)-based objective function (termed FCN(S) in short) has received increasing attention due to its simple structure and effectiveness of restoring clean speech signals from noisy counterparts. With evidence showing the benefits of FCN(S) for normal speech, this study sets out to assess its ability to improve the intelligibility of EAS simulated speech. Objective evaluations and listening tests were conducted to examine the performance of FCN(S) in improving the speech intelligibility of normal and vocoded speech in noisy environments. The experimental results show that, compared with the traditional minimum-mean square-error SE method and the deep denoising autoencoder SE method, FCN(S) can obtain better gain in the speech intelligibility for normal as well as vocoded speech. This study, being the first to evaluate deep learning SE approaches for EAS, confirms that FCN(S) is an effective SE approach that may potentially be integrated into an EAS processor to benefit users in noisy environments.

## I. INTRODUCTION

Cochlear implant (CI) is a surgically implanted electronic medical device that stimulates nerves to provide a sense of sound for people with profound-to-severe hearing loss. Despite technological and surgical advances since the 1960s, improving the speech perception and intelligibility of CI users in real-world scenarios remains challenging [1–5]. One promising direction is the combined electric and acoustic stimulation (EAS) technology. For EAS, an electrode array is implanted only partially into the cochlea because many people with hearing loss still have the residual acoustic hearing (20-60 dB hearing loss up to 750 Hz) at the low frequencies. That is, the device is a combination of hearing aid, which acoustically amplifies the low frequency signals, and a CI, which stimulates the regions responsible for the mid and high frequency sounds (for reviews on EAS fitting and signal processing, see [6, 7]). Dorman and Gifford [8] revealed that, compared to acoustic-only hearing aid and conventional CI users, EAS users achieved better speech recognition at both word and sentence levels. Although the benefits of EAS have been documented [9, 10], there is room for improvement in the performance of EAS in noisy environments. However, there is very little work on speech enhancement (SE) for EAS. Motivated by the advantage of EAS and the need for an effective SE approach for EAS devices, this study therefore explores whether deep-learning-based SE models are suitable for EAS, in comparison to a conventional SE approach. Moreover, this study examines whether SE approaches that have been found effective for CI are equally effective for EAS.

Currently, various SE models have been developed to cope with different noisy conditions [2,11–17]. These SE models are primarily used for conditions with a single microphone or multiple microphones. Compared to single-microphone approaches, multi-microphone approaches deal with spatially separated target and noise more efficiently [18–21]. Beamforming is constantly used to improve the recognition accuracy of multi-microphone speech data. For instance, Buechner et al. [22], compared omnidirectional microphone setting with two types of beamforming (adaptive monaural and binaural). The results demonstrated that both beamforming types yielded better speech perception scores than the omnidirectional approach. Despite satisfactory speech intelligibility can be achieved, multi-microphone approaches have some limitations. On the one hand, these approaches involve more hardware, such as a secondary microphone and headphone combination, and are therefore more expensive than the single-microphone methods. On the other hand, the applicability of multi-microphone approaches is restricted to the acoustic situation where the target and noise are spatially separated, and its efficacy degrades in reverberant environments [23]. Moreover, the speech signals acquired by multiple microphones are eventually fused to form a single-channel speech signal before being sent to EAS or CI users. Therefore, the effectiveness of a single-microphone SE approach plays an important role in the performance of EAS or CI devices.

Various single-microphone SE approaches have been proposed, which can be roughly divided into unsupervised and supervised approaches. A class of unsupervised SE approaches are derived based on the spectral and statistical properties of noise and speech signals; well-known approaches include spectral subtraction [24], minimum-mean square-error (MMSE) [25], logMMSE [26], Wiener-filter-based [27, 28], and signal-to-noise-ratio (SNR)-based [29, 30] methods. Another class of unsupervised SE approaches are the subspace-based methods, which construct two subspaces, one for clean speech and the other for noise signals, and use the information in the clean-speech subspace to restore the clean speech. Notable subspace techniques include singular value decomposition (SVD) [31], Karhunen–Loeve transform (KLT) [32, 33], and principal component analysis (PCA) [34]. Although many of these unsupervised single-microphone ap-proaches can produce satisfactory SE results for CI processors, they are more effective for stationary noise than for nonstationary noise, which does not always satisfy the unpredictable reality of acoustic conditions [35].

In addition to unsupervised SE approaches, numerous machine-learning-based algorithms have been popularly used in the single-channel SE field. For these approaches, a denoising model is usually prepared in a data-driven manner without imposing strong statistical assumptions on the clean speech and noise signals, and the noisy speech signal is processed by the denoising model to extract the clean speech signal. Notable examples include nonnegative matrix factorization [36], compressive sensing [37], and sparse coding [38]. More recently, deep learning models have been applied to the single-channel SE field. With deep structures, the complex correlation of noise and clean speech signals can be characterized. Deep-learning-based methods have demonstrated notable improvements over traditional methods [39]. Well-known deep-learning-based models include deep denoising autoencoder (DDAE) [15, 40], deep fully connected networks [35, 41, 42], recurrent neural networks (RNNs) [43–45], and convolutional neural networks (CNNs) [46, 47].

Recently, two research directions of deep-learning-based SE have attracted great attention. The first intends to develop a more appropriate input-output, and the second aims to derive a task-oriented objective function to train the denoising model. Aiming to identify suitable input-output, most conventional single-microphone approaches, such as DDAE-based SE, use power spectrum (PS) or its logarithmic form [14, 15, 40] as the acoustic features as the input of the denoising model. The denoising model aims to transform noisy PS features to generate enhanced features that are as close as possible to the clean references. To restore the speech waveform, the phase information from the original noisy speech is typically used. This is because there is no clear structure in a phase spectrogram, and it is difficult to accurately estimate the clean phase information from its noisy counterpart [46, 48]. It is clear that directly using the phase of the noisy speech is not optimal and may degrade the enhanced speech quality. Many approaches have been proposed to overcome this imperfect phase estimation issue and can be roughly divided into two categories. The first category adopts complex spectra as the acoustic features. The deep learning model learns the mapping or masking function to retrieve the clean complex spectrum from the noisy one, and thus simultaneously estimate the phase and amplitude information of the speech signal. Many studies have confirmed that complex spectral features lead to better performance than (log) PS features [49, 50]. The second category suggests that enhancement can be performed directly on a raw speech waveform without transforming it into spectral features [51–55]. For instance, Fu et al. [51] proposed using a fully convolutional neural network (FCN) model for SE in the time domain because FCN can preserve neighbouring information of a speech waveform to generate high and low frequency components using the same denoising model. Their experimental results show that, compared to CNN and deep neural networks (DNNs), the FCN model yields better speech intelligibility in terms of the short-time objective intelligibility (STOI) with fewer parameters. Later on, an utterance-based SE approach based on the FCN model was proposed in [52]. This utterance-based FCN model is capable of handling different kinds of objective functions from a local time scale (frame) to a global time scale (utterance) and achieves higher perceptual scores (STOI and perceptual estimation of speech quality, PESQ) than the frame-based FCN model.

The second focus of recent deep-learning-based SE research is to derive appropriate objective functions in consideration of human auditory perception. Conventional deep-learning-based SE employs engineering-defined distances, such as the mean square error (MSE) based on the L2 norm Euclidian distance and the L1 norm, to measure the error of the enhanced speech signal and the reference clean speech signal [15, 35, 40]. More recently, other evaluation metrics, such as PESQ [56, 57] and automatic speech recognition accuracy [58–61], have been adopted to form the objective functions for training the deep-learning models. Fu et al. [52] proposed to train the FCN model using the STOI-based objective function (termed FCN(S) in short) and observed that the enhanced speech signal is superior to speech signal generated by an MSE-trained FCN model in terms of STOI. Moreover, the results of subjective recognition tests performed on people with normal hearing (NH) also confirmed that the STOI-based objective function enables the FCN model to generate speech signals with higher intelligibility.

A recent study has evaluated the efficacy of the DDAE model for CI using a noise-vocoded speech simulation [14]. Experimental results show that the DDAE-based method outperforms three commonly used single-microphone SE approaches (logMMSE, KLT, and Wiener filter), in terms of intelligibility, evaluated with STOI, and speech recognition, evaluated with listening tests. The results have confirmed the potential of applying deep learning models to improve CI devices.

This study aims to evaluate the performance of FCN(S) with vocoded speech, which simulates speech signal processing used in EAS under various noisy conditions. The SE performance of FCN(S) is compared with a traditional SE approach, MMSE, and a DDAE model [14]. We tested the performance on both stationary and nonstationary noise types at two different SNR levels. Experimental results have confirmed that FCN(S) can achieve better performance in both objective evaluation and subjective listening tests compared to MMSE and DDAE.

The reminder of this paper is organized as follows: Section II introduces the architecture of the FCN(S) model. Section III presents the vocoder that is used to generate the vocoded speech signals. Section IV reports experimental setup and results as well as provides discussions about the results. Section V provides the concluding remarks of this study.

## II. THE FCN(S) MODEL

The FCN(S) model used in this study is proposed by Fu et al. [52]. Fig.1 shows the structure of the utterance-based FCN SE model. The FCN model can accommodate speech input with any arbitrary length. In the FCN model of Fig.1, each (1D CNN) filter is convolved with all the generated waveforms from the previous layer and produces filtered waveforms. Therefore, the filters have another dimension in the channel axis. Since the goal of the single-channel SE approaches is to generate one clean utterance, there is only one filter in the last layer. Note that the FCN SE model presented in Fig.1 is a complete end-to-end (noisy waveform in and clean waveform out) framework that does not require pre- and post-processing (feature extraction and speech restoration).

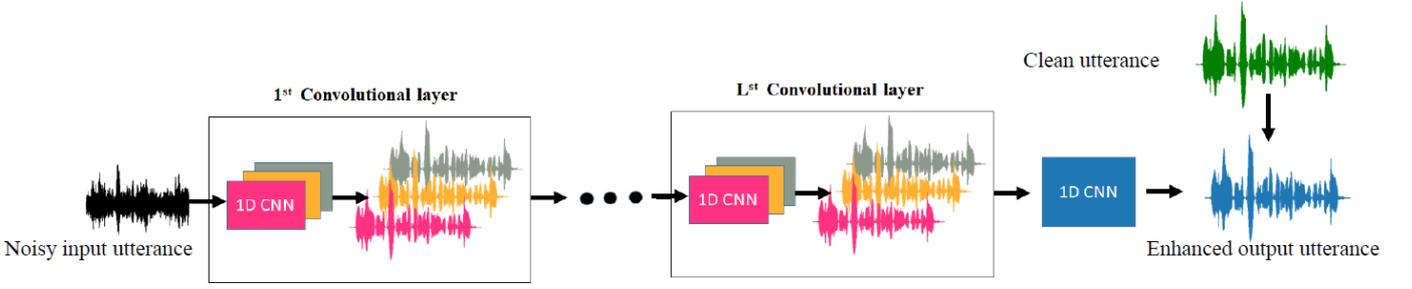

Fig.1. Structure of an utterance-based FCN SE model.

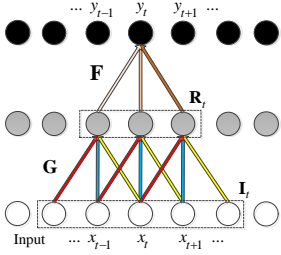

Fig. 2. The relation between the output layer and the last hidden layer in a fully convolutional framework.

The FCN model does not contain fully connected layers, as shown in Fig. 2. It is similar to the conventional CNN but all the fully connected layers are removed. Thus, the total number of parameters in FCN is considerably reduced. More importantly, in CNN with fully connected layers, the local information and the spatial arrangement of the previous layer could not be well preserved. By changing this design, the FCN model is capable of dealing with the high and low frequency components of the raw waveform at the same time. The relation between the output sample $y_t$ and the connected hidden nodes $\mathbf{R}_t$ can be represented by the following equation.

$$y_t = \mathbf{F}^T \mathbf{R}_t \quad (1)$$

where $\mathbf{F} \in \mathbb{R}^{f \times 1}$ denotes one of the learned filters, and $f$ is the size of the filter. Note that $\mathbf{F}$ is shared in the convolution operation and is fixed for each output sample. Therefore, if $y_t$ is in the high frequency region, $\mathbf{R}_t$ and $(\mathbf{R}_{t-1}, \mathbf{R}_{t+1})$ should be different. The similarity between $\mathbf{R}_t$ and its neighbors depends on the filtered outputs of previous locally connected nodes $\mathbf{I}_t$. For the details on the structure of the FCN model for waveform enhancement, please refer to the previous works [46, 47].

When we use the L2 norm, the objective function is defined as:

$$\mathcal{L}(\theta) = \sum_{t=1}^{T} \|y_t - q_t\|^2 \quad (2)$$

where $\theta$ denotes the model parameters of FCN, $y_t$ and $q_t$ are the $t$-th samples of the estimated and reference clean waveforms, respectively, and $T$ denotes the number of samples in the waveform.

When using STOI in the objective function, we have

$$\mathcal{L}(\theta) = -\frac{1}{U}\sum_u stoi(\mathbf{w}_y(u), \mathbf{w}_q(u)) \quad (3)$$

where $\mathbf{w}_y(u)$ and $\mathbf{w}_q(u)$ are the $u$-th estimated utterance and clean reference, respectively, and $U$ is the total number of training utterances. $stoi(.)$ is the function that calculates the STOI value of the noisy/processed utterance given the clean reference.

There are five steps to calculate the STOI value [62]:
1) Removing silent frames: Silent regions do not contribute to speech intelligibility and are therefore removed prior to evaluation.
2) Short-time Fourier transform (STFT): STFT is applied on both clean and noisy/processed speech utterances to obtain a representation that is similar to the speech properties in the auditory system.
3) One-third octave band analysis: 15 one-third octave bands are used to transform the clean and noisy/processed speech spectra with the lowest center frequency set to 150 Hz and the center-frequency of the highest one-third octave band set to 4.3 kHz.
4) Normalization and clipping: The respective goal of the normalization and clipping procedures is to compensate for global level differences and to ensure that the sensitivity of the STOI assessment to a severely degraded TF-unit is upper bounded.
5) Intelligibility measure: The intermediate intelligibility measure is defined as the correlation coefficient between the temporal envelopes of clean and noisy/processed speech signals. Finally, the STOI score is calculated as the average of the intermediate intelligibility measures on all bands and frames.

Notably, short segments (e.g., 30 frames) of temporal envelopes of the clean and the noisy/processed speech are used to compute the correlation coefficient. Therefore, the objective function of Eq. (3) cannot be directly optimized by a traditional frame-wise enhancement scheme. On the contrary, the FCN model that can take input of any arbitrary length can be combined appropriately with the STOI-based objective function. The FCN model optimized with the STOI-based objective function is termed FCN(S).

In [52], a disadvantage of using Eq. (3) as the objective function has been noted: the enhanced speech signals still involve clear noise components. To improve the noise suppression capability of FCN(S), we have derived a modified objective function that combines the MSE and STOI terms, which is represented as,

$$\mathcal{L}(\theta) = \frac{1}{U}\sum_u \left(\frac{\alpha}{L_u}\|\mathbf{w}_y(u) - \mathbf{w}_q(u)\|_2^2 - stoi(\mathbf{w}_y(u), \mathbf{w}_q(u))\right) \quad (4)$$

where $L_u$ is the length of the $u$-th utterances $\mathbf{w}_y(u)$ and $\mathbf{w}_q(u)$, and $\alpha$ is the weighting factor of the two optimization targets. The first term $(\frac{1}{L_u}\|\mathbf{w}_y(u) - \mathbf{w}_q(u)\|_2^2)$ in Eq. (4) denotes the sample-wise MSE. The combined objective function is to minimize the reconstruction errors while maximizing the STOI score.

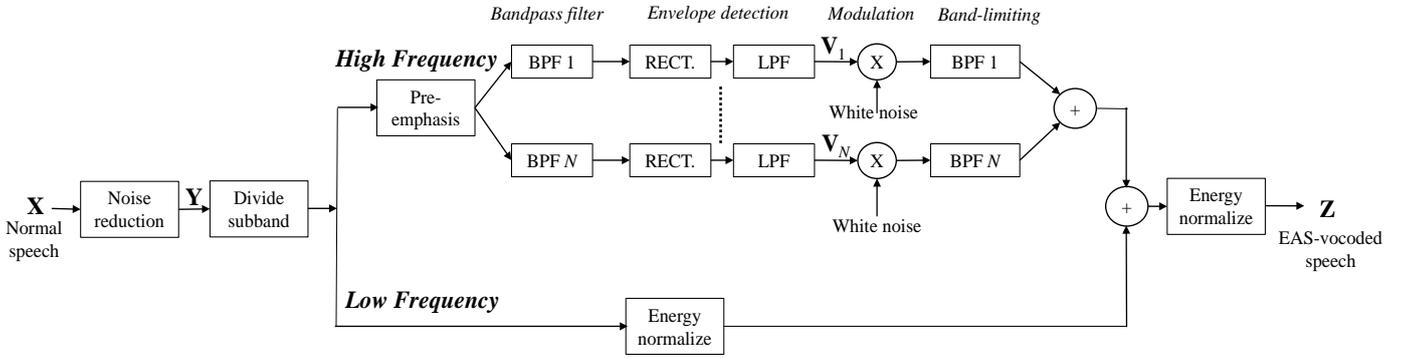

Fig. 3. Structure of the vocoder to simulate EAS speech. The system can be divided into electric and acoustic path. The electric path (the upper side) is formed by an *N*-channel noise-vocoder, and the acoustic path (the lower path) is the low-pass filtered speech signals.

## III. VOCODED SPEECH

Vocoder (Voice Operated reCOrdER) is a voice processing system for analysing and resynthesizing human voice signals [1, 58]. Vocoder has been widely used for audio data compression, voice encryption and transmission, and voice modification. In addition, vocoder has a profound impact on CI research. Using vocoder, speech signals are processed to simulate the sound heard by CI users, and the simulations are presented to NH participants for listening tests for various purposes, such as predicting the general pattern of speech recognition performance of CI users [14, 63–65]. Research using vocoder simulations may solve patient-recruitment issues as well as avoid patient-specific confounding factors, such as neural surviving patterns [64, 66]. Noise-vocoded speech, in particular, has been used in many studies to simulate CI speech processing [14, 67, 68] and produce reliable results. With some modifications to a CI vocoder, an EAS vocoder can be realized. In previous studies, the EAS vocoder has been used to simulate EAS speech in order to evaluate various perspectives of performance. One is to compare the performance between EAS and conventional CI devices [8, 69]. Another is to examine the influence of different coding parameters of EAS on speech intelligibility and recognition performance [70]. The other, the same as this study, adopts the EAS vocoder to evaluate speech recognition under noisy conditions [71, 72].

In this study, we adopted an EAS vocoder with the structured shown in Fig. 3. The normal speech signals, **X**, first passes through an SE stage to obtain an enhanced speech signal, **Y**. The enhanced signal is then processed through two paths, namely the acoustic and electric paths. For the acoustic path, the speech signal is directly processed by a low-pass filter (LPF). In this study, we used a Butterworth low-pass filter with a cut-off frequency of 500 Hz. The electric path is based on a standard CI noise vocoder that consists of pre-emphasis, band-pass filtering, envelope detection, modulation, and band-limiting stages. In the pre-emphasis stage, a 3dB/octave roll-off filter with a cut-off frequency of 2000 Hz is applied. In this study, four band-pass filters were used, i.e., *N*=4. With 4 band-pass filters, the emphasized signal is then divided into 4 frequency with cut-off frequencies of 500, 1017, 1901, and 3414 Hz. A full-wave rectifier is used to extract the 4-bands of temporal envelopes $\mathbf{V}_n$ (*n*=1,..,4) before the signals undergo a low-pass filter with a cut-off frequency of 400 Hz. The envelopes for all bands are then modulated with a set of white noise before further filtered by the same set of band-pass filters. Finally, the modulated sinewaves of the four bands and the filtered acoustic signal are summed, and the level of the combined signal is adjusted to produce a root-mean-square value equal to the original input wideband signal. Finally, we obtained the EAS-vocoded speech (Z in Fig. 3).

## IV. EXPERIMENTAL SETUP AND RESULTS

In this study, the evaluation focused on the ability of FCN(S) to improve speech intelligibility under training-testing mismatched conditions. More specifically, the training and testing utterances were prepared using different scripts, recorded by different speakers, and contaminated by different noise types. The speech corpus for evaluation consisted of 2,560 Mandarin utterances recorded by 8 native speakers (4 males and 4 females), each of whom recorded 320 utterances. The recording scripts were the Taiwan Mandarin hearing in noise test (TMHINT) [73]. Each sentence contained 10 Chinese characters, and the corresponding speech length was about 3-4 seconds. All utterances were recorded in a quiet environment with a sampling rate of 16 kHz. We selected the first 200 utterances of 6 speakers (3 males and 3 females) as the clean training data. The last 120 utterances of the remaining 2 speakers (1 male and 1 female) were used to prepare the testing set. One hundred noise types from a database of 100 non-speech environmental sounds [74] were adopted to corrupt clean training utterances to produce signals with SNR from -10 to 20dB. We randomly selected 30,000 utterances for the training set. The engine and street noises (different from those used in the training set) were used to generate -3 and 1dB signals for the testing set. The performance of traditional MMSE and DDAE was tested for comparison.

Two sets of experiments were conducted, including the evaluation of normal speech and the evaluation of vocoded speech. For normal speech, the STOI scores for normal wideband speech processed by the three SE approaches were reported. In many SE studies, the STOI score has been used as a standardized evaluation metric to measure the speech intelligibility. The range of the STOI score is from 0 to 1; the higher the STOI value, the better the speech intelligibility. As mentioned earlier, we also performed a listening test using an EAS vocoder. In the test, the utterances processed by the three SE approaches and the EAS vocoder were presented to the normal hearing participants. The speech recognition results for each SE approach were then measured based on the listening test.

## A. Evaluation on Normal Speech

The spectrograms of the two noise types, namely engine and street, are presented in Fig. 4. The spectrogram plot can show how the frequency patterns present in a sequential signal changes over time [75]. From the spectrograms, different characteristics of the engine (as stationary) and street (as non-stationary) noises can be easily observed. In the following, we first provide qualitative comparisons of different SE approaches using waveform and spectrogram plots. Then, we present the results of quantitative evaluation using the STOI scores.

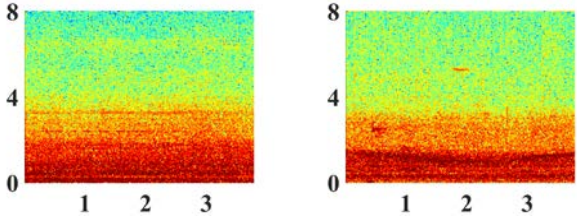

Fig. 4. Spectrogram plots of noise signals: (a) engine noise and (b) street noise. The two noise types were used to synthesize the noisy speech of the testing data. In both plots, the horizontal axis is the time in second, and the vertical axis is the frequency in kHz.

### 1) Waveform and Spectrogram Analyses on Normal Speech

Waveform and spectrogram plots are commonly used to visually observe the characteristics of a time-varying signal series. The waveform plot can directly display the sample values along the time index and provide complimentary information to the spectrogram plot. Fig. 5 (c), (d), and (e) show the waveform and spectrogram plots of enhanced speech by the MMSE, DDAE, and FCN(S) SE methods under street noise at -3dB SNR, respectively. For comparison, the spectrogram and waveform plots for the clean and noisy speech signals are shown in Fig. 5(a) and (b), respectively. For a fair comparison, all the speech signals in Fig. 5 are normalized to zero-mean and unit-variance.

By comparing the spectrogram plots, we observe that the three SE approaches exhibit different denoising characteristics. As shown in the spectrogram of Fig. 5(c), MMSE removed some of the high frequency components of the street noise signal but the mid- to low- frequency speech signal was considerably distorted. More specifically, when the speech and noise fell in the mid-low frequency region (please see the dashed box (1) in Fig. 5(c)), the MMSE showed limited noise reduction capability. Moreover, some speech components were removed (please see the dashed box (2) in Fig. 5(c)). On the other hand, although DDAE (cf. Fig. 5(d)) effectively removed the high frequency noise components, it overly-removed speech structures in the mid-low frequency region. More specifically, the mid-low frequency speech components that overlapped with the same frequency band of the noise (shown in the dashed box in Fig. 5(d)) were misjudged as noise and were removed. As for the FCN(S) method (cf. Fig. 5(e)), although some noise components remained, it maintained a clearer speech structure in the mid-low frequency region than MMSE and DDAE. The weakness of FCN(S) seemed to be dealing with the high frequency regions where the noise component was more notable than the DDAE result (please see the dashed box in Fig. 5(e)). Yet, it has been pointed out that the mid-low frequency region carries more speech intelligibility information [76], so it is more important to accurately restore this region when the goal is to maximize speech intelligibility. This argument has been discussed and confirmed experimentally in our previous study [52].

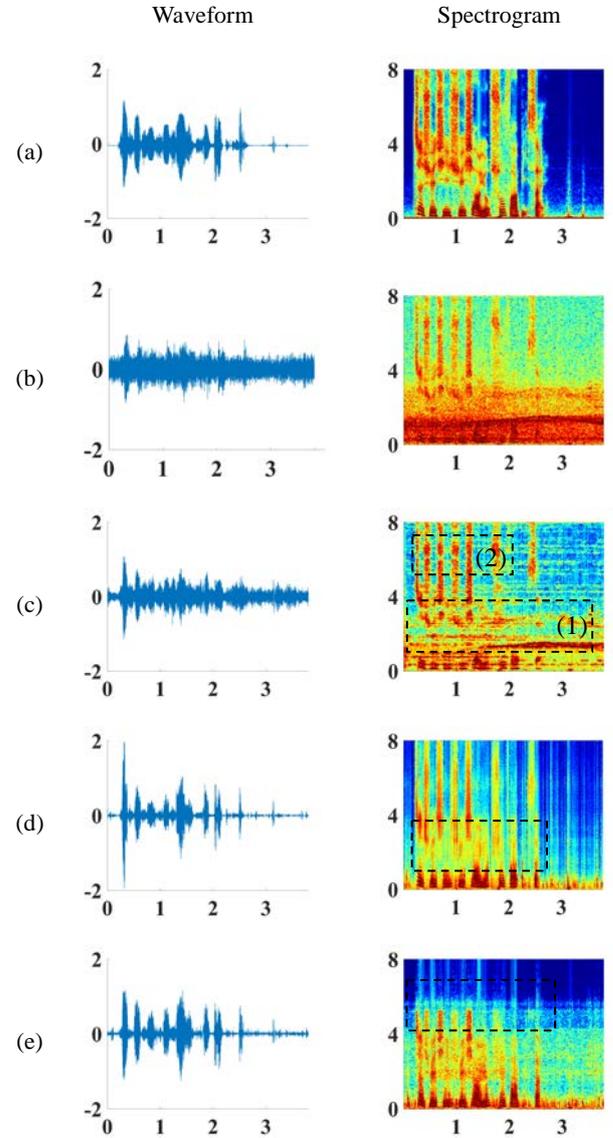

Fig. 5. Waveform and spectrogram plots of a testing utterance: (a) clean speech, (b) noisy speech (street noise; SNR at -3dB), enhanced speech by (c) MMSE, (d) DDAE, and (e) FCN(S). In the waveform and spectrogram plots, the horizontal axis is the time in second. In the waveform plots, the vertical axis is the normalized value; while in the spectrogram plots, the vertical axis is the frequency in kHz.

### 2) Objective Evaluation on Normal Speech

To objectively evaluate the SE performance, we tested the STOI scores for the three SE approaches. The average STOI scores at six different SNR levels for engine and street noise types are demonstrated in Fig. 6(a) and (b), respectively. The results of unprocessed speech are denoted as "Noisy" in the following presentation. For the results of the engine noise, as shown in Fig. 6(a), the average STOI scores for {Noisy, MMSE, DDAE, FCN(S)} are {0.15, 0.17, 0.30, 0.31} at -11dB, {0.25, 0.28, 0.44, 0.49} at -7dB, {0.38, 0.41, 0.59, 0.64} at -3dB, {0.51, 0.54, 0.71, 0.73} at 1dB, {0.64, 0.67, 0.78, 0.79} at 5dB, and {0.73, 0.80, 0.81, 0.83} at 9dB. For the results of the street noise, as shown in Fig. 6(b), the average STOI scores for {Noisy, MMSE, DDAE, FCN(S)} are {0.20, 0.22, 0.37, 0.42} at -11dB, {0.32, 0.33, 0.54, 0.60} at -7dB, {0.44, 0.45, 0.66, 0.71} at -3dB, {0.55, 0.55, 0.75, 0.77} at 1dB, {0.65, 0.67, 0.80, 0.81} at 5dB, and {0.74, 0.77, 0.82, 0.85} at 9dB.

The results in Fig. 6 demonstrate that FCN(S) achieved better STOI scores across different SNR levels than the other two SE approaches, regardless of the noise type. As shown in Fig. 6(a) and (b), although the conventional SE approach,

MMSE, achieved slightly better STOI scores than unprocessed noisy speech for stationary noise, it is generally ineffective for nonstationary noise. Compared to MMSE, both DDAE and the FCN(S) demonstrated notable improvement margins (with very similar patterns) across the six different SNR levels. However, it is evident that in lower SNR levels (-7dB, -3dB, and 1dB SNRs) FCN(S) outperformed DDAE. The results in Fig. 6 therefore suggest that FCN(S) can provide better speech intelligibility than DDAE under more challenging SNR conditions for both stationary and non-stationary noise types.

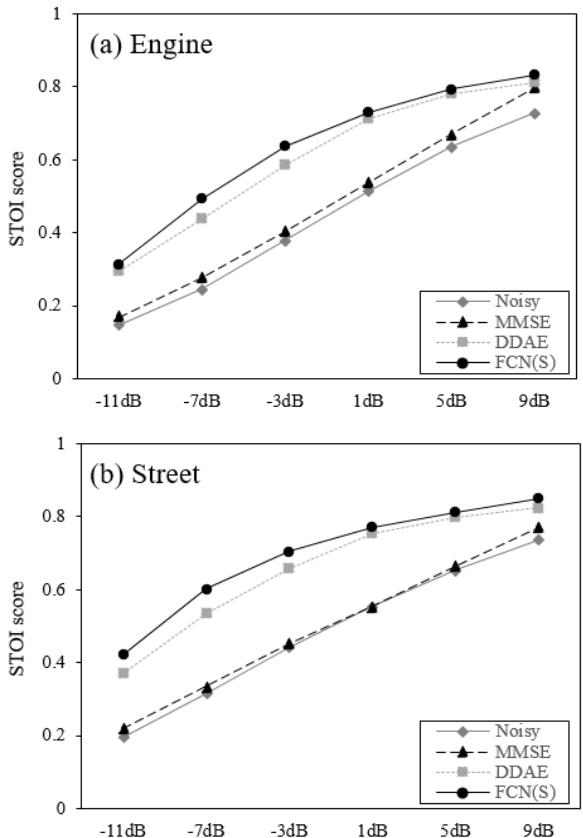

Fig. 6. Average STOI scores at seven SNR levels for (a) engine and (b) street noises.

### B. Evaluation on Vocoded Speech

Next, we investigate the performance of the three SE approaches on EAS vocoded speech. Similarly, we first visually compare the amplitude envelop and spectrogram plots of the three SE approaches. Then, we present the recognition results of the subjective listening test.

*1) Amplitude Envelop and Spectrogram Analyses on Vocoded Speech*

In addition to waveform and spectrogram plots, amplitude envelop plots are another useful tool for analyzing time-varying signals [77, 78]. Previous studies [63, 77, 78] have reported a positive correlation of modulation depth (of amplitude envelop) and speech intelligibility. It has also been shown that the middle frequency band is more important for speech intelligibility than the low- and high-frequency bands. Therefore, we examined the amplitude envelop of the second channel of the speech signal processed by the three different SE approaches. Fig. 7(c), (d), and (e), respectively, show the amplitude envelop and spectrogram plots of enhanced speech by MMSE, DDAE, and FCN(S) under street noise at -3dB SNR. For comparison, the amplitude envelop and spectrogram plots of the clean and noisy speech signals are shown in Fig. 7(a) and (b), respectively. The same as Fig. 5, we have nor-malized all the speech signals reported in Fig. 7 to be zero-mean and unit-variance for a clear comparison.

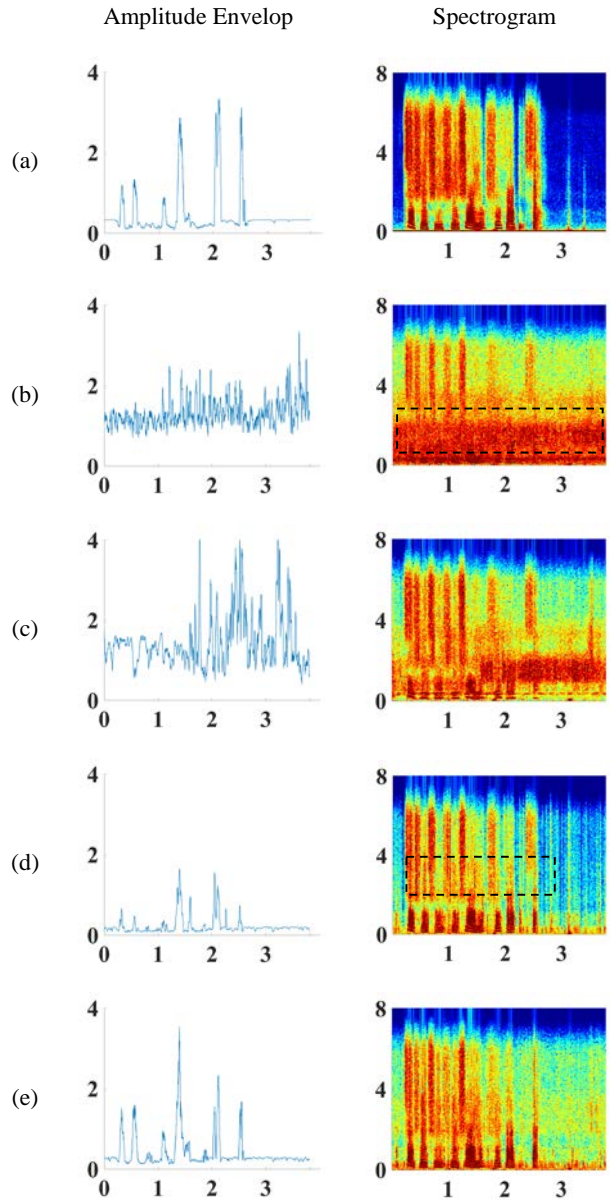

Fig. 7. Spectrograms of a vocoded speech utterance: (a) clean speech, (b) noisy speech (street noise; SNR at -3dB), (c) MMSE enhancement, (d) DDAE enhancement, and (e) FCN(S) enhancement. In the waveform and spectrogram plots, the horizontal axis is for time in second. In the amplitude envelop plots, the vertical axis is for normalized value; in the spectrogram plots, the vertical axis is for frequency in kHz.

The amplitude envelop plots provide another qualitative comparison of the three SE approaches. The modulation depth is associated with speech perception accuracy. As shown in Fig. 7(c), MMSE caused the loss of amplitude information, resulting in poor speech intelligibility. DDAE showed good SE performance; the envelop profile was clear but the modulation depth was slightly sacrificed. FCN(S) outperformed both MMSE and DDAE, showing a higher modulation depth, indicating better speech intelligibility.

When comparing the spectrogram plots in Fig. 7, we note that the main trends in vocoded speech simulation are largely in line with those in normal speech (in Fig. 5). MMSE could not effectively reduce the noise components, as highlighted in the dotted box in Fig. 7(c). DDAE, as shown in Fig. 7(d), retained most of the speech structure, although some speech components were clearly overly-removed. Finally, FCN(S) preserved the speech structure well while effectively reducing the noise components, comparing the spectrogram plot in Fig. 7(e) with the spectrogram plots of the clean and noisy speech signals in Fig. 7(a) and (b). In the previous normal speech evaluation, FCN(S) showed weakness in dealing with high frequency noise. However, as shown in the dotted box in Fig. 7(e), the speech features in the high frequency region were less affected when using FCN(S), compared to the other two SE models. Based on the spectrograms in Fig. 7, we can predict that FCN(S) will provide better speech intelligibility, which is discussed in the next section.

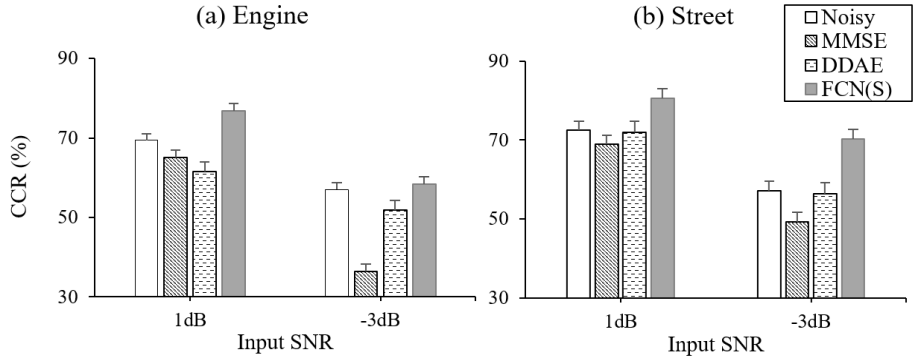

Fig. 8. Average speech recognition scores in terms of character correct rate (%) in 1 and -3dB SNR condition for (a) engine and (b) street noise masks. The error bars indicate one standard error of the mean (SEM).

*2) Subjective Listening Test on Vocoded Speech*

The performances of the SE approaches (MMSE, DDAE, and FCN(S)) on EAS vocoded speech at two SNR levels (-3dB and 1dB) were evaluated by native Mandarin Chinese speakers, aged 18-39, with NH. To avoid listening fatigue, each participant experienced only one SNR level. Two groups of thirty participants, fifteen males and fifteen females, were recruited for the test. As with the objective evaluation, the test adopted the utterances in the TMHINT dataset; engine and street noise types were used to corrupt the utterances. That is, in the listening test, a participant was presented with eight test conditions: 1 SNR level (-1 or 3dB) × 2 noise types (street and engine) × 4 SE models (Noisy, MMSE, DDAE, and FCN(S)). Each condition contained ten utterances, consisted of ten Chinese characters. The original noisy speech (termed Noisy) was used as baseline for evaluation. Although increasing evidence that both MMSE and DDAE can improve regular CI devices [2, 14, 71], whether the findings are replicable in the case of SE for EAS device requires verification.

The listening test was conducted in a quiet room with a set of Sennheiser HD headphones at a comfortable listening level. The participants were instructed to listen to each utterance and verbally repeat the sentence as completely as possible. An utterance could be repeated once before the participants gave their response. Overall, 80 utterances were played to the participants during the test. The presentation order of the test conditions was randomized, and none of the sentence was repeated across the test conditions. The listening test took about forty-five minutes to complete. During the test, the participants were allowed to have a short break whenever they requested. For the evaluation metric, the character correct rate (CCR) was calculated by dividing the number of correctly recognized characters by the total number of characters under each test condition. As in this test, the total number of characters per test condition is one hundred; therefore, the CCR is equal to the raw score of accuracy.

The average CCRs for all test conditions are demonstrated in Fig. 8. For the engine noise type, as shown in Fig. 8(a), the average±SEM CCRs for {Noisy, MMSE, DDAE, FCN(S)} are {69.8±1.7, 65.2±1.3, 61.6±1.3, 76.9±1.7} at 1dB SNR and {56.7±2.5, 36.8±2.7, 51.8±2.1, 58.4±2.5} at -3dB.

For the street noise type, as shown in Fig. 8(b), the average± SEM CCRs for {Noisy, MMSE, DDAE, FCN(S)} are {72.5± 1.7, 68.9±2.1, 72.0±2.2, 80.6±1.4} at 1dB SNR and {57.2±2.3, 49.4±2.3, 56.5±2.1, 70.3±2.5} at -3dB.

From the results in Fig. 8, we can note that FCN(S) consistently improved speech recognition across different noise types and SNRs, in comparison to baseline (Noisy). Moreover, the FCN(S) achieves higher CCRs than the DDAE and the MMSE for both stationary (engine) and nonstationary (street) noise. It is also noted that the other deep-learning-based SE approach, the DDAE, performed lower than baseline and similar to the performance of MMSE. Also, both the DDAE and the MMSE showed slightly worse performance on engine noise than on street noise. Although the performance of the DDAE on engine noise condition was below baseline, its performance on street noise was at the same level as baseline. This indicates that the DDAE is more effective for dealing with nonstationary noise than stationary noise under a challenging noise condition. At low SNR level, the performance of the MMSE was significantly worse than both deep-learning-based SE approaches, especially in the stationary noise condition.

To further verify the effectiveness of FCN(S) over Noisy and the other two SE approaches, we conducted dependent t-Test for matched-pair samples (one-tailed). The test dependent t-Test is particularly suitable for evaluating the difference between the test conditions here. For the test, we consider $H_0$ as "approach two is not better than approach one", and $H_1$ as "approach two is better than approach one". the CCR performance of the SE approaches under three test conditions (engine and street noise types with 1 and -3dB SNRs). The results of the dependent T-tests for 1dB SNR level is demonstrated in Table 1. From the table, it is clear that FCN(S) notably outperforms Noisy, DDAE, and MMSE for both test conditions, with very small P-values (much smaller than 0.005).

Table 1. P-values of the dependent T-tests for test conditions at 1dB SNR level.

| SE approach 1 vs. SE approach 2 | | P-value | |
|---|---|---|---|
| | | Engine | Street |
| FCN(S) | Noisy | 2.526E-04 | 1.474E-04 |
| | DDAE | 1.647E-06 | 2.610E-04 |
| | MMSE | 1.690E-05 | 7.200E-06 |

On the other hand, the results of the dependent T-tests for -3dB SNR level are presented in Table 2. Similar to the patterns observed for the 1dB SNR level (in Table 1), the FCN(S) significantly improved speech recognition over Noisy, DDAE, and MMSE (with very small P-values) for both engine and street noise types.

Table 2. P-values of the dependent T-tests for test conditions at -3dB SNR level.

| SE approach 1 vs. SE approach 2 | | P-value | |
|---|---|---|---|
| | | Engine | Street |
| FCN(S) | Noisy | 2.891E-01 | 1.686E-06 |
| | DDAE | 1.562E-03 | 1.724E-04 |
| | MMSE | 3.252E-10 | 1.210E-08 |

The results of the listening tests for EAS vocoded speech are mostly in line with the STOI evaluation. These findings further reveal that the conventional SE approach, the MMSE, cannot effectively reduce noise components and thus yield the lowest recognition scores. Although the DDAE outperformed the MMSE in most test conditions, the CCRs in stationary environment at 1dB was nearly the same as that of the MMSE. This result reflects the advantage of the DDAE for nonstationary noise at low SNRs and the advantage is diluted in stationary noise environment. Finally, the FCN(S) effectively enhanced speech utterance across SNRs and noise type.

The benefit of the FCN(S) may come from two perspectives. One is the input and output formats of the denoising model, and the other is the STOI-based objective function. By directly enhancing the speech signals in the waveform domain, the FCN(S) reduces the distortions caused by imperfect phase estimation. Moreover, an utterance-based optimization is adopted to train the FCN model. As reported in [52], the utterance-based optimization can more effectively characterize the temporal correlations of long speech segments in order to optimize perception-based objective function. In [14], it is reported that the DDAE can effectively reduce noise components from the noisy speech signals and accordingly improve the speech intelligibility of CI-vocoded speech. The same effect was not observed for EAS-vocoded speech in this study. It should be noted that in [14], the training and testing utterances are prepared by the same male speaker, and thus the SE system is actually speaker-dependent. In the present study, we designed a more challenging and realistic scenario, where the trainig and testing utterances are prepared by different speakers. The better intelligiblity achieved by FCN(S) suggest that the model has better generalization capability to training-testing mismatched conditions. Further, the FCN(S) actually considers both the STOI and the MSE (as shown in Eq. (4)) in model optimization so that it could maintain a good balance between speech intelligibility and quality.

## V. CONCLUSION

This study is the first to investigate the effectiveness of deep learning based speech enhancement methods on EAS simulated speech. We focus on comparing the recently developed FCN(S) SE approach with a conventional MMSE SE approach and a deep-learning-based DDAE SE approach (proven to provide promising performance in CI simulation) at two different SNRs in stationary and nonstationary noisy environments. Based on the results of objective evaluation and listening tests, FCN(S) outperformed the other two SE approaches under all test conditions. The findings of this study confirm the advantages of FCN(S) for SE in EAS and demonstrate that it is a more effective SE model than MMSE and DDAE in several perspectives. First, FCN(S) outperforms MMSE and DDAE regardless of noise type. That is, the FCN(S) model deals with both stationary noise (engine) and nonstationary noise (street) better than the other two SE models. Second, the effectiveness of FCN(S) is particularly evident under more challenging SNR conditions, e.g., -3dB. Third, with fewer parameters than DDAE, FCN(S) provides more gains than the other two SE models across test conditions. Overall, the main contribution of this study is to verify the efficacy of FCN(S) for enhancing normal as well as EAS-vocoded speech. Based on the current evidence, FCN(S) can be an effective choice to be implemented in EAS speech processors to increase speech intelligibility.